\documentclass[twocolumn,showpacs,preprintnumbers,amsmath,amssymb]{revtex4}
\usepackage{epsfig}
\usepackage{graphicx}
\usepackage{dcolumn}
\usepackage{bm}

\usepackage{amsthm}
\usepackage{amsfonts}
\usepackage{color} 

\newcommand{\ket}[1]{\mbox{$ | #1 \rangle $}}
\newcommand{\bra}[1]{\mbox{$ \langle #1 | $}}

\newcommand{\ud}{\mathrm{d}}

\begin{document}

\title{Passive sources for the Bennett-Brassard 1984 quantum key distribution protocol with practical signals}
\author{Marcos Curty$^1$, Xiongfeng Ma$^2$, Hoi-Kwong Lo$^3$, and
Norbert L\"utkenhaus$^{2}$}
\affiliation{$^1$ ETSI Telecomunicaci\'on, Department of Signal Theory and Communications, University of Vigo, 
Campus Universitario, E-36310 Vigo, Pontevedra, Spain \\
$^2$ Institute for Quantum Computing \& Department of Physics and Astronomy, University of Waterloo, N2L 3G1 
Waterloo, Ontario, Canada \\
$^3$ Center for Quantum Information and Quantum Control, Department of Physics and Department of 
Electrical \& Computer Engineering, University of Toronto, M5S 3G4 Toronto, Ontario, Canada}

\begin{abstract}
Most experimental realizations of quantum key distribution are based on the Bennett-Brassard 1984 
(so-called BB84) protocol. In a typical optical implementation of this scheme, the sender uses an active source
to produce the required BB84 signal states. 
While active state preparation of BB84 signals
is a simple and elegant solution in principle, in 
practice passive state preparation might be desirable in some scenarios, for instance, 
in those experimental setups operating at high transmission rates. Passive schemes 
might also be more robust against side-channel attacks than active sources. 
Typical passive devices involve parametric down-conversion. 
In this paper, we show that both coherent light and practical single photon sources 
are also suitable for passive generation of BB84 signal states. Our method 
does not require any external-driven element, but only 
linear optical components and photodetectors. In the case of coherent light, the resulting key rate 
is similar to the one delivered by an active source. When the sender uses practical 
single photon sources, however, 
the distance covered by a passive transmitter might be longer than the one of 
an active configuration.
\end{abstract}

\maketitle

\section{Introduction}

The transmission of secret information over an insecure 
communication channel constitutes an essential resource in modern information society. 
Quantum key distribution (QKD) is a technique that 
allows two distant parties
(typically called Alice and Bob)
to establish a secure secret key despite the computational and technological power 
of an eavesdropper (Eve), who interferes with the signals \cite{qkd}. This secret key is the main
ingredient of the one-time-pad or Vernam cipher \cite{Vernam}, 
the only known encryption method that can deliver information-theoretic secure communications. 

Most experimental realizations of QKD are based on the so-called BB84 QKD
scheme introduced by Bennett and Brassard in 1984 \cite{bb84}. In a typical quantum 
optical implementation of this protocol, Alice sends to Bob phase-randomized weak coherent pulses (WCP)
with usual average photon number of 0.1 or higher \cite{4lasers, modulator}. 
These states can be easily generated using only standard semiconductor lasers and calibrated attenuators. 
Each light pulse may be prepared in a different polarization state, which is selected, 
independently and randomly for each signal, between 
two mutually unbiased bases, {\it e.g.}, either a linear (H [horizontal] or V [vertical]) or 
a circular (L [left] or R [right]) polarizations basis. 
On the receiving side, Bob measures each incoming signal by choosing at random 
between two polarization analyzers, one for each
 possible basis. 
Once this quantum communication phase is completed, Alice and Bob use an authenticated public 
channel to process their data and obtain a secure secret key. 
This last procedure, called key distillation, involves, typically, 
local randomization, 
error correction to reconcile Alice's and Bob's data, and privacy ampliÞcation to 
decouple their data from Eve \cite{post}. A full proof of the security for the BB84 QKD protocol with WCP
has been given in Refs.~\cite{sec_bb84,sec_bb84b}.
The performance of this scheme can be improved further if the original hardware is 
slightly modified. For instance, one can use the so-called decoy-state method \cite{decoy,decoy_e}, 
where Alice varies the mean photon number 
of each signal state she sends to Bob. The measurement results associated to different 
intensity settings allow the legitimate 
users to obtain a better estimation of the behavior of the quantum channel. This  
translates into an enhancement of the achievable secret key rate, which can now basically 
reach the performance of single photon sources (SPS). 

\begin{figure}
\begin{center}
\includegraphics[angle=0,scale=0.6]{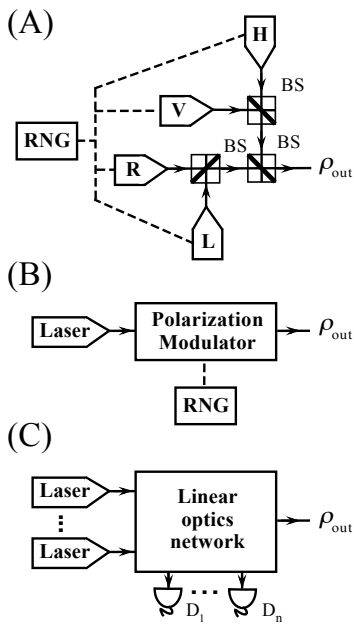}
\end{center}
\caption{(Case A) Example of an active setup with four laser diodes controlled by a random 
number generator (RNG). Each source emits one different BB84 signal state
({\it e.g.}, H [horizontal], V [vertical], L [circular left] or R [circular right] polarization states). 
BS denotes a beamsplitter.
(Case B) Example of an active setup with one laser diode together 
with a polarization modulator controlled 
by a RNG. This modulator rotates the state of polarization of the incoming pulses 
to generate BB84 signal states. (Case C)
Basic setup of a passive QKD transmitter with linear optics elements. One or more laser diodes produce 
different signal states that are sent through a linear optics network. Depending on 
the detection pattern observed in the detectors $D_i$, different signal states are actually 
generated. No RNG or active modulator are necessary in this last scenario. 
\label{figure_intro}}
\end{figure}

The preparation of the BB84 signal states is usually realized by means of an active source. 
For simplicity, we will first consider polarization encoding. (Note that phase encoding is mathematically equivalent 
to polarization encoding. Later in this paper, we will also consider phase encoding.)
There are two main configurations; they are illustrated in Fig.~\ref{figure_intro} as cases A and B.
In the first one (case A in the figure), Alice uses 
four laser diodes, one for each possible BB84 signal \cite{4lasers}. These lasers are controlled 
by a random number generator (RNG) \cite{random} that decides each given time
which one of the 4 diodes is triggered. 
The second configuration (case B in the figure) uses only one single laser diode 
in combination with a polarization modulator which is
controlled 
by a RNG \cite{modulator}. This modulator can rotate the state of polarization of the signals 
produced by the source. 

While active state preparation is a simple and elegant solution
to implement the BB84 protocol in principle, in practice 
passive state preparation might be desirable in some scenarios \cite{rarity,passive1,passive2}; for instance, 
in those experimental setups operating at high transmission rates, since no RNG is required in a passive device. A passive 
transmitter allows Alice to generate different quantum states at random without the need to use an external 
source of randomness. 
This situation is illustrated as case C in Fig.~\ref{figure_intro}. Alice can use 
one or more light sources to produce 
different signal states that are sent through a linear optics network. Depending on 
the detection pattern observed in her detectors $D_i$, she can infer which signal states are actually 
generated. So far, only entanglement-based (EB) QKD systems have been designed to operate entirely in a passive regime 
without any external-driven elements \cite{rarity,passiveEB}. For instance, Alice and Bob can 
use a beamsplitter (BS)
to passively and randomly select which bases to measure each incoming pulse.
In this article, we show that both coherent light and
practical SPS are also suitable for passive generation of 
BB84 signal states, {\it i.e.}, one does not 
need a nonlinear optics network preparing entangled states in order to passively 
generate BB84 signals. Practical SPS (also called ``sub-Poissonian sources'') are those 
light sources which have a smaller probability of emitting two or more photons
than an attenuated laser.
Our method requires only linear optical elements and 
photodetectors. In the 
asymptotic limit of an infinite long experiment, it turns out that the secret key rate of a 
passive source with coherent light is similar to the one delivered by an active source. 
When Alice uses 
practical SPS, however, the distance covered by a passive transmitter might be longer 
than the one of an active configuration.
This result is caused by the capacity of the passive scheme to reduce the 
multiphoton probability of the source via a post-selection mechanism.

Passive schemes might also 
be more robust than active systems to side-channel attacks hidden in the imperfections of the optical
components. 
If a polarization modulator is not properly designed, for example, it 
may distort some of the physical parameters of the pulses emitted 
by the sender depending on the particular value of the polarization setting selected. This fact could open a 
security loophole in the active schemes. 

The article is organized as follows. 
In Sec.~\ref{sec_one} we present and evaluate the performance of a 
passive BB84 state 
preparation scheme that can operate with coherent light.
Then, in Sec.~\ref{sec_two} we consider 
the case where Alice uses practical SPS. 
Finally, Sec.~\ref{conc} concludes the article with a summary. The paper 
includes as well a few appendixes with additional calculations.

\section{Passive transmitter with coherent light}\label{sec_one}

\begin{figure}
\begin{center}
\includegraphics[angle=0,scale=0.66]{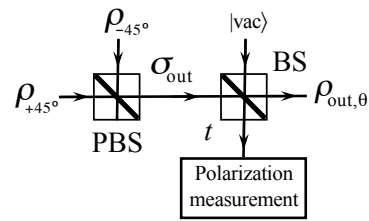}
\end{center}
\caption{Basic setup of a passive BB84 QKD source with strong coherent light. 
The mean photon number of the signal states $\rho_{+45^{\circ}}$ and
$\rho_{-45^{\circ}}$ can be chosen very high; for instance, $\approx{}10^8$ photons. The parameter 
$t$ represents the transmittance of the BS; it satisfies $t\ll{}1$.
\label{figure1}}
\end{figure}

The basic setup is rather simple. It is illustrated in Fig.~\ref{figure1}. 
Suppose two 
phase-randomized strong coherent pulses prepared, respectively, in $+45^{\circ}$ and $-45^{\circ}$ 
linear polarization, interfere at a polarizing beamsplitter (PBS). These states can be written as 
\begin{eqnarray}
\rho_{+45^{\circ}}&=&e^{-\frac{\upsilon}{2}}\sum_{n=0}^\infty\frac{(\upsilon/2)^n}{n!}\ket{n_{+45^{\circ}}}\bra{n_{+45^{\circ}}}, \nonumber \\
\rho_{-45^{\circ}}&=&e^{-\frac{\upsilon}{2}}\sum_{n=0}^\infty\frac{(\upsilon/2)^n}{n!}\ket{n_{-45^{\circ}}}\bra{n_{-45^{\circ}}},
\end{eqnarray}
where $\ket{n_{\pm45^{\circ}}}$ denote Fock states with $n$ photons in $\pm45^{\circ}$ linear polarization. 
The mean photon number, $\upsilon/2$, of each signal
can be chosen very high; for instance, $\approx{}10^8$ photons. In this scenario, we will show that
the signal $\sigma_{{\rm out}}$ at the output port of the PBS can be expressed as
\begin{equation}
\sigma_{{\rm out}}=\frac{1}{2\pi}e^{-\upsilon{}}\sum_{n=0}^\infty \frac{\upsilon{}^n}{n!}\int_{\theta} \ket{n_{\theta}}\bra{n_{\theta}}\ \ud \theta,
\end{equation}
where the Fock states $\ket{n_{\theta}}$ are given by
\begin{equation}\label{fock1}
\ket{n_{\theta}}=\frac{\Big[\frac{1}{\sqrt{2}}\big(a_{+45^{\circ}}^\dagger+e^{i\theta}a_{-45^{\circ}}^\dagger\big)\Big]^n}{\sqrt{n!}}\ket{vac}.
\end{equation}
Here $\ket{vac}$ denotes the vacuum state, and $a_{\pm45^{\circ}}^\dagger$
are the creation operators for the  $\pm45^{\circ}$ linear polarizations modes. 

To see this, let us consider the interference of two pure coherent states with fixed phase relation
and orthogonal polarizations, 
$\ket{\sqrt{\upsilon/2}e^{i\phi_1}}_{+45^{\circ}}$ and $\ket{\sqrt{\upsilon/2}e^{i\phi_2}}_{-45^{\circ}}$, at 
a PBS. The output signal is a coherent state of the form
\begin{equation}
\ket{\sqrt{\upsilon}e^{i\phi_2}}_{\theta}=e^{-\upsilon/2}\sum_{n=0}^\infty\frac{\big(\sqrt{\upsilon}e^{i\phi_2}\big)^n}{\sqrt{n!}}\ket{n_{\theta}},
\end{equation}
with $\theta=\phi_1-\phi_2$.
The case of two phase-randomized coherent pulses can be solved by just integrating 
the signal $\ket{\sqrt{\upsilon}e^{i\phi_2}}_\theta$ over all angles $\phi_2$ and $\theta$, {\it i.e.},
\begin{eqnarray}
\sigma_{{\rm out}}&=&\frac{1}{(2\pi)^2}\int_{\theta}\int_{\phi_2} \ket{\sqrt{\upsilon{}}e^{i\phi_2}}_{\theta}\bra{\sqrt{\upsilon{}}e^{i\phi_2}}\ \ud \phi_2 \ud \theta \nonumber \\
&=&\frac{1}{2\pi}e^{-\upsilon{}}\sum_{n=0}^\infty \frac{\upsilon{}^n}{n!}\int_{\theta} \ket{n_{\theta}}\bra{n_{\theta}}\ \ud \theta.
\end{eqnarray}

Now, to prepare the signal states that are sent to Bob, Alice performs a 
polarization measurement followed by a post-selection step. By assumption,
we have that the intensity $\upsilon$ 
of the signals $\sigma_{{\rm out}}$ is very high. Therefore,  
Alice can always employ, for instance, a BS of very small transmittance ($t\ll{}1$) to split 
these states into two light beams: one 
very weak suitable for QKD, and 
one strong. The weak signal is sent to Bob through the quantum channel
(see Fig.~\ref{figure1}). The strong beam is used to measure its polarization by means 
of a polarization measurement which, for simplicity, we assume is {\it perfect}. For each
incoming signal, this device provides Alice with a precise value for the measured angle $\theta$.
In this situation, the conditional states emitted by the source can be described as 
\begin{equation}\label{basicstate}
\rho_{{\rm out}, \theta}=e^{-\mu}\sum_{n=0}^\infty \frac{\mu^n}{n!} \ket{n_{\theta}}\bra{n_{\theta}},
\end{equation}
where $\theta$ denotes the value of the angle obtained by Alice's 
polarization measurement, and $\mu$ is given by $\mu=\upsilon{}t$.
In practice, however, as we will show below, it is sufficient if the polarization measurement
tells Alice the value of $\theta$ within a certain interval.

Whenever $\theta\in\{0, \pi/2, \pi, 3\pi/2\}$ Alice 
generates one of the four BB84 polarization states perfectly. 
Note that 
the creation operators for the polarizations modes in the BB84 protocol, which we shall denote as
$b_{\rm H}^\dagger$, $b_{\rm V}^\dagger$, $b_{\rm L}^\dagger$ and $b_{\rm R}^\dagger$, can be expressed, in terms 
of the operators $a_{\pm45^{\circ}}^\dagger$, as:
\begin{eqnarray}\label{fock2}
b_{\rm H}^\dagger&=&\frac{1}{\sqrt{2}}\big(a_{+45^{\circ}}^\dagger+a_{-45^{\circ}}^\dagger\big), \nonumber \\ 
b_{\rm V}^\dagger&=&\frac{1}{\sqrt{2}}\big(a_{+45^{\circ}}^\dagger-a_{-45^{\circ}}^\dagger\big), \nonumber \\
b_{\rm L}^\dagger&=&\frac{1}{\sqrt{2}}\big(a_{+45^{\circ}}^\dagger+ia_{-45^{\circ}}^\dagger\big), \nonumber \\
b_{\rm R}^\dagger&=&\frac{1}{\sqrt{2}}\big(a_{+45^{\circ}}^\dagger-ia_{-45^{\circ}}^\dagger\big). 
\end{eqnarray}

\begin{figure}
\begin{center}
\includegraphics[angle=0,scale=0.58]{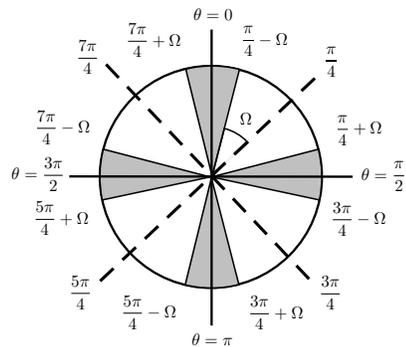}
\end{center}
\caption{Graphical representation of the valid regions for the angle $\theta$. These regions are marked 
in grey. They depend on an acceptance parameter $\Omega\in[0,\pi/4]$.
\label{figure1_bloch}}
\end{figure}

In general, Alice does not need to restrict herself to only those events where
she actually prepares a perfect BB84 state, since the probability 
associated with these ideal events tends to zero. Instead, she can also 
accept signals with a polarization sufficiently 
close to the desired ones. 
This situation is illustrated in Fig.~\ref{figure1_bloch}, 
where Alice selects some valid regions for the angle $\theta$.
These regions are marked 
with grey color
in the figure. They depend on an acceptance parameter $\Omega\in[0,\pi/4]$ that we optimize. 
Specifically, whenever the value of $\theta$ lies within any of the intervals 
$\psi\pm(\pi/4-\Omega)$ with $\psi\in\{0, \pi/2, \pi, 3\pi/2\}$, then 
Alice considers the pulse emitted by the source as a valid signal. 
This last condition can also be written as
\begin{equation}
\theta\in\bigcup_{1=0}^3\Big[(2i+1)\frac{\pi}{4}+\Omega,(2i+3)\frac{\pi}{4}-\Omega\Big].
\end{equation}
Otherwise, the pulse is discarded afterwards during the post-processing 
phase of the protocol, and it does not contribute to the key rate. 
The probability that a pulse is accepted, $p_{\rm acc}$, is given by
\begin{equation}
p_{\rm acc}=\frac{8\big(\frac{\pi}{4}-\Omega\big)}{2\pi}=1-\frac{4\Omega}{\pi}.
\end{equation}
To increase this probability, one can reduce the value of $\Omega$. Note, however, that this action 
also results in an increase of the quantum bit error rate (QBER) of the protocol, 
that we shall denote as $E$. On the other hand, when $\Omega$ increases, we have that both 
$p_{\rm acc}$ and $E$ decrease. There is a trade-off on the acceptance parameter $\Omega$. A high acceptance probability 
$p_{\rm acc}$ favors $\Omega\approx{}0$, whereas a low QBER favors $\Omega\approx{}\pi/4$. 
Note that in the limit where 
$\Omega$ tends to $\pi/4$ we recover the standard BB84 protocol.

\subsection{Lower bound on the secret key rate}\label{lower}

The single photon signals emitted by the 
passive source presented above, averaged over the values of Alice's 
key bit, are basis-independent. That is, they 
do not leak any information to Eve about the basis of Alice's signal states. 
To see this, note that
\begin{eqnarray}
&&\int_{\frac{7\pi}{4}+\Omega}^{\frac{\pi}{4}-\Omega} \ket{1_{\theta}}\bra{1_{\theta}}\ \ud \theta+
\int_{\frac{3\pi}{4}+\Omega}^{\frac{5\pi}{4}-\Omega} \ket{1_{\theta}}\bra{1_{\theta}}\ \ud \theta  \nonumber \\
&=&\int_{\frac{\pi}{4}+\Omega}^{\frac{3\pi}{4}-\Omega} \ket{1_{\theta}}\bra{1_{\theta}}\ \ud \theta+
\int_{\frac{5\pi}{4}+\Omega}^{\frac{7\pi}{4}-\Omega} \ket{1_{\theta}}\bra{1_{\theta}}\ \ud \theta \nonumber \\
&=&\frac{(\pi-4\Omega)}{4}\openone,
\end{eqnarray}
for all $\Omega\in[0,\pi/4]$, and where the single photon state 
$\ket{1_{\theta}}$ is given by Eq.~(\ref{fock1}). Here 
$\openone$ denotes the identity operator in the single
photon subspace. 

To evaluate the performance of this passive source  
we use the security analysis 
provided by Gottesman-Lo-L\"utkenhaus-Preskill (GLLP) in Ref.~\cite{sec_bb84b}. 
See also Ref.~\cite{koashi_p}. It considers that Eve can always obtain full information about 
the part of the key generated from the multiphoton signals. 
This pessimistic assumption is also true for the passive transmitter illustrated in Fig.~\ref{figure1}
when Alice and Bob use only unidirectional classical communication during the 
public-discussion phase of the protocol. This result arises from the fact that 
all the photons contained in a pulse are prepared in the same polarization state 
and, therefore, no secret key can be distilled with one-way post-processing techniques \cite{oneway}. 
Note, however, that such security analysis could leave still room for improvement when 
Alice and Bob employ two-way classical communication. In this situation,   
it might be possible to obtain secret key even from the multiphoton pulses since the signal states 
prepared by the passive device are already mixed at the source. This last scenario, however, 
is beyond the scope of this paper.

We further assume the typical initial post-processing step in the BB84 protocol, where double click 
events are not discarded by Bob, 
but they are randomly assigned to single click events \cite{squash1,squash2}. 
The secret key 
rate formula can be written as \cite{sec_bb84b}
\begin{eqnarray}\label{keyrateWCP}
R&\geq&{}qp_{\rm acc}\Big\{(Q-p_{\rm multi})\big[1-H(E_1)\big]\nonumber \\
&-&Qf(E)H(E)\Big\}.
\end{eqnarray}
The parameter $q$ is the efficiency of the protocol ($q=1/2$ for the standard BB84 protocol, and 
$q\approx{}1$ for its efficient version \cite{eff_bb84}); $Q$ is the gain, {\it i.e.}, the probability that 
Bob obtains a click in his measurement apparatus when Alice sends him a signal state;
$f(E)$ is the efficiency of the error 
correction protocol as a function of the error rate $E$ \cite{eff_error}, typically $f(E)\geq{}1$ with 
Shannon limit $f(E)=1$; $H(x)$ is the binary Shannon entropy function
defined as $H(x)=-x\log_2{(x)}-(1-x)\log_2{(1-x)}$;
$p_{\rm multi}$ is the multiphoton probability of the source, {\it i.e.},
\begin{equation}
p_{\rm multi}=1-e^{-\mu}(1+\mu);
\end{equation}
and $E_1$ denotes an upper bound on the single photon error rate. In the case 
of the standard BB84 protocol without decoy-states, this last quantity is given 
by \cite{sec_bb84b}
\begin{equation}
E_1=\frac{E}{1-\frac{p_{\rm multi}}{Q}}.
\end{equation}

For simulation purposes we use the channel model and detection device on Bob's side described 
in Appendix~\ref{channel}. This model allows us to calculate the observed experimental 
parameters $Q$ and $E$. These quantities are given in Appendix~\ref{QE_coherent}.
Our results, however, can also be applied to any other quantum 
channel or detection setup, as they depend only on the observed gain and QBER. 

\begin{figure}
\begin{center}
\includegraphics[angle=0,scale=0.34]{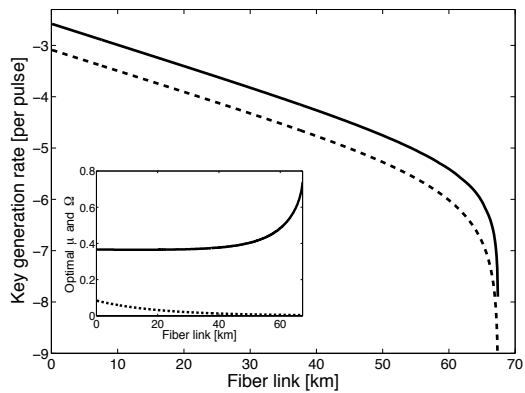}
\end{center}
\caption{
Lower bound on the secret key rate $R$ given by Eq.~(\ref{keyrateWCP}) 
in logarithmic scale 
for the passive source illustrated in Fig.~\ref{figure1} (dashed line). 
The signal states $\rho_{{\rm out}, \theta}$ are given by Eq.~(\ref{basicstate}).
For simulation purposes, we consider the following experimental parameters: the dark count rate 
of Bob's detectors is
$\epsilon_{\rm B}=1\times10^{-6}$, the overall 
transmittance of Bob's detection apparatus is $\eta_{\rm B}=0.1$, the loss coefficient of the 
channel is $\alpha=0.2$ dB/km, $q=1/2$, and the efficiency of the error correction protocol is $f(E)=1.22$.
The solid line
represents a lower bound on $R$ when Alice employs an active source. 
The inset figure shows the value for the optimized parameters 
$\mu$ (dashed line) and 
$\Omega$ (solid line) 
 in the passive setup.  
\label{figure2}}
\end{figure}

The resulting lower bound on the secret key rate is illustrated in Fig.~\ref{figure2} (dashed line). 
In our simulation we employ the following experimental parameters:
the dark count rate of Bob's detectors is
$\epsilon_{\rm B}=1\times10^{-6}$, the overall 
transmittance of his detection apparatus is $\eta_{\rm B}=0.1$, and the loss coefficient of the 
quantum channel is $\alpha=0.2$ dB/km.
We further assume that $q=1/2$, and $f(E)=1.22$.
These data are used as well for the simulations included in Sec.~\ref{sec_two}.
With this configuration, it turns out that the optimal value
of the mean photon number $\mu$ decreases with the distance, while the value of the parameter 
$\Omega$ increases. In particular, $\mu$ diminishes from $\approx{}0.084$ to 
approximately
$4\times{}10^{-3}$, while $\Omega$ augments from $\approx{}0.365$ to 
$\approx{}0.76$.
This result is not surprising. At long distances the gain $Q$ of the protocol is very low  
and, therefore, it is especially important to keep both the multiphoton probability 
of the source, and  
the intrinsic error rate of the signals $\rho_{{\rm out}, \theta}$, also low. Fig.~\ref{figure2} includes  
an inset plot with the optimized parameters 
$\mu$ (dashed line) and 
$\Omega$ (solid line). This figure shows as well a lower bound 
on the secret key rate for the case of an active source. 
The cutoff point where the secret key rate drops down to zero 
is basically the same in
both cases. It is given by
$l\approx{}67.5$ km. This result arises from two main 
limiting factors:
the multiphoton probability of the source, and the dark count rate of Bob's detectors. 
Note that in these simulations we do not consider any misalignment effect in the 
channel or in Bob's detection apparatus. 
From the results 
shown in Fig.~\ref{figure2} 
we see that the performance of the passive scheme is similar 
to the one of an active setup, thus showing the practical interest of the passive source. 
The (relatively small) difference between the achievable secret key rates  
in both scenarios is due to two main factors: (a) the probability $p_{\rm acc}$ to 
accept a 
pulse emitted by the source, which is $p_{\rm acc}<1$ in the passive setup, and $p_{\rm acc}=1$ in 
the active scheme, and (b) the intrinsic error rate of the signals accepted by Alice, 
that is zero only in the case of an active source.

\subsection{Alternative implementation scheme}\label{alter}

\begin{figure}
\begin{center}
\includegraphics[angle=0,scale=0.66]{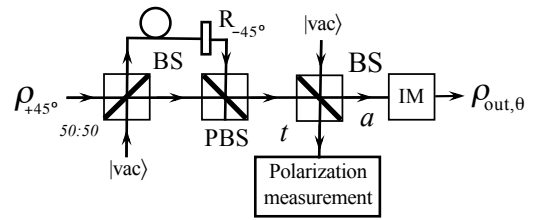}
\end{center}
\caption{Alternative implementation scheme with only one pulsed laser source. 
The delay introduced by one arm of the interferometer is equal to the time difference 
between two consecutive pulses. The polarization rotator $R_{-45^{\circ}}$ changes the 
$+45^{\circ}$ linear polarization of the incoming pulses to $-45^{\circ}$ 
linear polarization. The intensity modulator (IM) blocks either all the even or all the odd 
optical pulses in mode $a$.
\label{figure3}}
\end{figure}

The passive setup shown in Fig.~\ref{figure1} requires that Alice employs two independent 
sources of phase-randomized strong coherent pulses. Alternatively to this scheme, Alice could as well 
employ, for instance, 
the device
illustrated in Fig.~\ref{figure3}.
This setup has
only one laser diode, but follows a similar spirit like the original scheme shown in 
Fig.~\ref{figure1}, where a polarization measurement is used to determine the 
polarization of the incoming signals. The main idea is just to replace two single light pulses emitted by two 
different diodes with two consecutive light pulses generated by only one laser diode. 

To keep the analysis simple, the scheme includes as well  
an intensity modulator (IM) to block either all the even or all the odd pulses in mode $a$
(see Fig.~\ref{figure3}). The main reason for blocking half of these pulses is to suppress possible correlations between 
them. That is, the action of the IM  guarantees that the signals that go to Bob are
precisely 
tensor product of states of the form given by Eq.~(\ref{basicstate}). This way  
we can directly apply the security 
evaluation provided in the previous section. This transmitter requires, therefore, an active control of the functioning of the IM. 
Note, however,  
that this configuration might still be much less of a problem than using a polarization modulator
to actively generate BB84 signal states at high rates, since no RNG 
is needed to control the IM. 
Thanks to the one-pulse delay introduced by one arm of the interferometer,
it can be shown that both setups in Fig.~\ref{figure1} and
Fig.~\ref{figure3} are completely equivalent, except 
from the resulting secret key rate. More precisely, 
the secret key rate
in the passive 
scheme with two lasers is 
double than that in the setup illustrated in Fig.~\ref{figure3}, 
since half of
the pulses are now discarded. 

\begin{figure}
\begin{center}
\includegraphics[angle=0,scale=0.66]{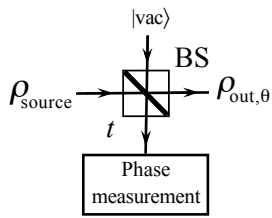}
\end{center}
\caption{Basic setup of a passive BB84 QKD phase encoding transmitter with strong 
coherent light. The mean photon number of the signal states $\rho_{\rm source}$ 
can be chosen very high; for instance, $\approx{}10^8$ photons. 
\label{figure_phase}}
\end{figure}

To conclude, let us mention that similar ideas to the ones presented in this 
section can also be used in other implementations of the BB84 protocol with a
different signal encoding. One example are those QKD experiments based on 
phase encoding, which turns out to be 
more suitable to use in combination with optical fibers than 
polarization encoding \cite{qkd}. This situation
is illustrated in Fig.~\ref{figure_phase}, where now Alice uses a BS of very small 
transmittance ($t\ll{}1$) to split phase-randomized 
strong coherent pulses into two light beams. 
The strong beam is 
used to measure the value of its phase relative to some local reference phase 
by means of a phase measurement, 
while Alice sends the weak signal to Bob.  
Like in the passive source shown in Fig.~\ref{figure1}, Alice can select some valid 
regions for the measured phase, and the analysis presented before also applies
straightforwardly 
to this scenario.

\section{Passive transmitter with single photon sources}\label{sec_two}

The passive state preparation schemes with coherent light introduced in the previous section
deliver a key generation rate of order $O(\eta_{\rm sys}^2)$, 
where $\eta_{\rm sys}$ denotes the overall transmittance of the system. 
To achieve higher secure key rates over longer distances there are two 
main alternatives: To combine them with the decoy-state method \cite{decoy,decoy_e}, 
or to employ SPS. In the first case, Alice can vary the mean photon number 
of the BB84 signals just by using a 
variable optical attenuator or a passive decoy-state setup \cite{passive2}. 
Here, we will concentrate on the second scenario, and we present 
a simple passive BB84 source which uses practical
SPS. Note that the passive transmitters analyzed in Sec.~\ref{sec_one} cannot be employed
with SPS, since those setups need to operate with light pulses 
of sufficiently high intensity to be able to measure their polarization 
with certain precision.

\begin{figure}
\begin{center}
\includegraphics[angle=0,scale=0.66]{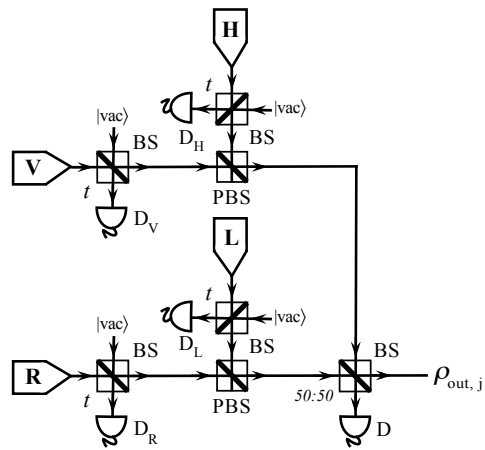}
\end{center}
\caption{Basic setup of a passive BB84 QKD transmitter with practical SPS. Each 
laser diode emits pulses prepared in a different BB84 polarization state: 
H [horizontal], V [vertical] linear polarizations, and L [left], R [right] circular polarizations. 
The parameter $t$ represents the transmittance of the BS's. All detectors shown in the 
figure  
denote threshold single photon detectors.  
\label{figure4}}
\end{figure}

The basic setup is illustrated in 
Fig.~\ref{figure4}.
It contains four light sources, each of them preparing a different BB84 polarization state.
The working principle of this device is rather simple. Let us first consider the ideal case. 
That is, each photon source in the figure 
emits precisely one single photon in the desired polarization state, 
and all detectors have perfect detection efficiency 
($\eta_{\rm det}=1$) and dark count rate $\epsilon_{\rm A}$ equal to zero. 
In this scenario, we have that 
whenever Alice observes a click in precisely three of the ``signal detectors" 
$D_{\rm H}$, $D_{\rm V}$, $D_{\rm L}$,
and $D_{\rm R}$, and no click in detector $D$ (see Fig.~\ref{figure4}), the state 
generated by the source, that we shall denote as $\rho_{{\rm out},j}$ with $j\in\{{\rm H, V, L, R}\}$, 
consists of just 
one photon prepared in the polarization state associated with the signal detector $D_j$ which 
did not click. For example, suppose that Alice obtains a click in detectors $D_{\rm V}$, $D_{\rm L}$,
and $D_{\rm R}$, and no click in detectors $D_{\rm H}$ and $D$, 
then the output state is given by $\rho_{\rm out,H}=\ket{1,0}_{l}\bra{1,0}$, 
where $\ket{1,0}_{l}$ denotes a Fock state with one 
photon in the horizontal polarization mode and zero photons 
in the vertical polarization mode.    

In order to evaluate the performance of this setup in a more realistic scenario,  
we shall consider practical SPS emitting Fock diagonal 
states of the form
\begin{equation}\label{rhosource}
\rho_{{\rm source},j}=\sum_{n=0}^\infty p_n \ket{n}_j\bra{n},
\end{equation}
with $j\in\{{\rm H, V, L, R}\}$. Furthermore, for simplicity, 
we assume that the photon number distribution $p_n$
of each source is the same for all them, independently of their polarization.  

To characterize the threshold 
single photon 
detectors $D_{\rm H}$, $D_{\rm V}$, $D_{\rm L}$, $D_{\rm R}$, and $D$,
we use a POVM 
with two elements, $F_{\rm{vac}}$ and $F_{\rm click}$, given by
\begin{eqnarray}\label{rome_final}
F_{\rm{vac}}&=&(1-\epsilon_{\rm A})\sum_{n=0}^{\infty} (1-\eta_{\rm det})^n \ket{n}\bra{n},
\end{eqnarray}
and $F_{\rm click}=\openone-F_{\rm{vac}}$, where again the parameter $\eta_{\rm det}$
denotes the detection efficiency of the 
detector, and $\epsilon_{\rm A}$ represents its probability of having a dark count. 
We
assume that $\epsilon_{\rm A}$ is, to a good 
approximation, independent of the incoming signals.
The outcome of $F_{\rm{vac}}$ corresponds to no click in the detector, while the operator 
$F_{\rm click}$ gives precisely one detection click, which means at least one photon is 
detected.
 
After a tedious calculation, one can obtain the conditional output state $\rho_{{\rm out},j}$
given that Alice 
observes a click only in three of the signal detectors
$D_{i}$ and no click in detectors $D$ and $D_j$ (with $j\neq{}i$). A mathematical expression for 
this state, together with its associated probability, is 
given in Appendix~\ref{cond_state}. 

\subsection{Lower bound on the secret key rate}

Like in Sec.~\ref{lower}, the single photon signals emitted by the passive source 
illustrated in
Fig.~\ref{figure4} are also basis-independent.
To analyze the performance of this source we use again the security results 
provided by GLLP in Ref.~\cite{sec_bb84b}. See also Ref.~\cite{koashi_p}.
The secret key rate formula is given by Eq.~(\ref{keyrateWCP}). 
Note, however, 
that such security analysis might overestimate the eavesdropping capabilities of Eve
in this scenario even when Alice and Bob use one-way post-processing techniques
during the public-discussion phase of the protocol. In contrast to the case analyzed 
in Sec.~\ref{sec_one}, now multiphoton pulses can contain photons prepared in different 
polarization states. 
This effect is similar to the one observed in those EB implementations
of the BB84 protocol that use, for instance,
a pulsed type-II down conversion source \cite{pump}.
In this sense, the passive source shown in Fig.~\ref{figure4} 
might be more robust against 
the Photon Number Splitting (PNS) attack 
\cite{pns} than an 
active one, since now Eve might not be able to  
obtain always full information about the part 
of the key generated with the multiphoton signals.
Still, the multiphoton problem is now on Bob's side who gets a noisy signal,
which can contain photons not prepared in Alice's state.
The possibility to distill a secret key from multiphoton pulses 
in this situation constitutes an interesting theoretical question that 
deserves further investigations, but it is 
beyond the scope of this paper.

\begin{figure}
\begin{center}
\includegraphics[angle=0,scale=0.34]{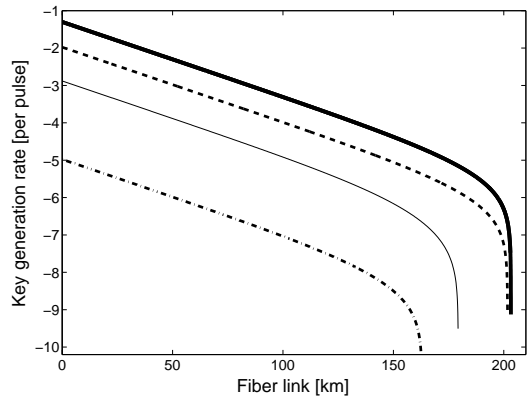}
\end{center}
\caption{
Lower bound on the secret key rate $R$ given by Eq.~(\ref{keyrateWCP}) 
in logarithmic scale 
for the passive setup illustrated in Fig.~\ref{figure4} with perfect on-demand 
SPS, 
{\it i.e.}, $p_1=1$ in Eq.~(\ref{rhosource}). 
In the simulation we consider the following experimental parameters: the dark count rate 
of Alice's and Bob's detectors is
$\epsilon_{\rm A}=\epsilon_{\rm B}=1\times10^{-6}$, the overall 
transmittance of Bob's detection apparatus is $\eta_{\rm B}=0.1$, the loss coefficient of the 
channel is $\alpha=0.2$ dB/km, $q=1/2$, and the efficiency of the error correction protocol is $f(E)=1.22$.
We further assume the channel model described in Appendix~\ref{channel}, where we neglect any 
misalignment effect. Otherwise, the actual secure distance will
be smaller. 
The figure includes three cases, depending on the actual value of the efficiency 
$\eta_{\rm A}$ of Alice's detectors, 
and we optimize the transmittance $t$ of the BS's on Alice's side for each case. In particular, 
we assume
$\eta_{\rm A}=1$ (dashed line), $\eta_{\rm A}=0.5$ (thin solid line), and 
$\eta_{\rm A}=0.1$ (dash-dotted line).
The solid line represents a lower bound on $R$ when Alice employs an active source that emits 
single photon pulses with probability one.  
\label{figure6}}
\end{figure}

\begin{figure}
\begin{center}
\includegraphics[angle=0,scale=0.34]{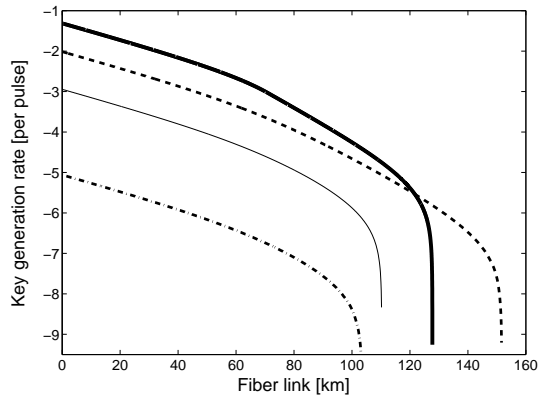}
\end{center}
\caption{
Lower bound on the secret key rate $R$ given by Eq.~(\ref{keyrateWCP}) 
in logarithmic scale 
for the passive setup illustrated in Fig.~\ref{figure4} with imperfect SPS.
In particular, we consider that the photon number distribution $p_n$ of the sources is given by
$p_0=0.0099$, $p_1=0.9882$, and $p_2=1-p_0-p_1=0.0019$.
We further assume the same experimental data used in Fig.~\ref{figure6}.
We study three cases, depending on the actual value of the efficiency 
$\eta_{\rm A}$ of Alice's detectors, 
and we optimize the transmittance $t$ of the BS's on Alice's side for each case. Specifically, 
we assume
$\eta_{\rm A}=1$ (dashed line), $\eta_{\rm A}=0.5$ (thin solid line), and 
$\eta_{\rm A}=0.1$ (dash-dotted line).
The solid line represents a lower bound on $R$ when Alice employs an active source
with photon number distribution $p_n$ in combination with a BS. In this last case, 
we optimize the value of the transmittance of the BS as a function of the distance.   
\label{figure5}}
\end{figure}

\begin{figure}
\begin{center}
\includegraphics[angle=0,scale=0.33]{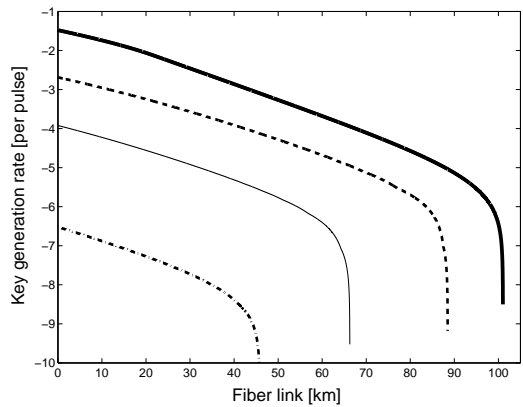}
\end{center}
\caption{
Lower bound on the secret key rate $R$ given by Eq.~(\ref{keyrateWCP}) 
in logarithmic scale 
for the passive setup illustrated in Fig.~\ref{figure4} with imperfect SPS.
The photon number distribution $p_n$ of the sources is given by
$p_0=0.2$, $p_1=0.785$, and $p_2=1-p_0-p_1=0.015$.
We further assume the same experimental data used in Fig.~\ref{figure6}.
We study three cases, depending on the actual value of the efficiency 
$\eta_{\rm A}$ of Alice's detectors, 
and we optimize the transmittance $t$ of the BS's on Alice's side for each case. In particular, 
we assume
$\eta_{\rm A}=1$ (dashed line), $\eta_{\rm A}=0.5$ (thin solid line), and 
$\eta_{\rm A}=0.1$ (dash-dotted line).
The solid line represents a lower bound on $R$ when Alice employs an active source
with photon number distribution $p_n$ in combination with a BS. In this last case, 
we optimize the value of the transmittance of the BS as a function of the distance.
\label{figure_6b}}
\end{figure}

To evaluate the secret key rate formula given by Eq.~(\ref{keyrateWCP}), we need 
to obtain the gain $Q$, the error rate $E$, and the multiphoton
probability $p_{\rm multi}$ of the source. These parameters are calculated in 
Appendix~\ref{sps_par}. For that, we use again the channel model and detection 
device described in Appendix~\ref{channel}. 
The resulting lower bound on $R$ is 
illustrated in Figs.~\ref{figure6}, \ref{figure5}, and \ref{figure_6b}, 
for three different photon number distributions $p_n$ of the practical SPS.
For simulation purposes, we use the same experimental data employed
in Sec.~\ref{lower}, and we further assume that Alice's detectors have 
a dark count rate $\epsilon_{\rm A}=1\times10^{-6}$. 

The first example 
plotted in Fig.~\ref{figure6}
considers the case of perfect on-demand SPS, {\it i.e.}, we impose 
$p_1=1$ in Eq.~(\ref{rhosource}). We study three different situations, 
depending on the actual value of the efficiency 
$\eta_{\rm A}$ of Alice's detectors, 
and we optimize the transmittance $t$ of the BS's on Alice's side for each case. 
In particular, we assume $\eta_{\rm A}=1$, $\eta_{\rm A}=0.5$, and 
$\eta_{\rm A}=0.1$. This figure includes as well a lower bound on $R$ when Alice employs an active source
that emits 
single photon pulses (thick solid line). 
The cutoff points where the secret key rate drops down to zero are given by
$l\approx{}201.7$ km (passive source with $\eta_{\rm A}=1$), 
$l\approx{}179.3$ km (passive source with $\eta_{\rm A}=0.5$), 
$l\approx{}162.5$ km (passive source with $\eta_{\rm A}=0.1$), and
$l\approx{}203$ km (active scheme).
From the results shown in this figure we 
see that, in this ideal scenario where $p_1=1$, 
the performance of the passive scheme is similar
to the one of an active setup only when the 
efficiency of Alice's detectors is relatively high. 
Note that the difference between the achievable secret key rates  
in both scenarios comes mainly from the probability $p_{\rm acc}<1$ that Alice's transmitter 
produces a valid signal in the passive scheme 
({\it i.e.}, three of her signals detectors click). 
Specifically, 
when the efficiency $\eta_{\rm A}$ of Alice's detectors 
is too low, then also the probability $p_{\rm acc}$  
decreases significantly, and therefore the secret key rate decreases as well.  
The value of $\eta_{\rm A}$ also influences the resulting cutoff points for the secret 
key rate. For instance, when $\eta_{\rm A}$ decreases, the intrinsic noise of Alice's signals 
({\it i.e.}, the probability to produce a wrong signal) can increase due to the dark counts of 
her detectors. Note that this effect becomes more relevant for low $\eta_{\rm A}$. 
As a result, the cutoff point for the secret key rate decreases. 

The other two examples illustrated in Figs.~\ref{figure5} and \ref{figure_6b}
analyze the influence that the vacuum and multiphoton probabilities 
of the practical SPS can have on the final secret key rate. For that, we consider 
two different photon number 
distributions $p_n$ for the SPS described in Eq.~(\ref{rhosource}). In particular, 
Fig.~\ref{figure5} shows the case where $p_0=0.0099$, $p_1=0.9882$, and $p_2=1-p_0-p_1=0.0019$,
while Fig.~\ref{figure_6b} assumes 
$p_0=0.2$, $p_1=0.785$, and $p_2=1-p_0-p_1=0.015$. In both cases, we evaluate 
the same three scenarios contemplated in Fig.~\ref{figure6}, {\it i.e.}, we consider
$\eta_{\rm A}=1$, $\eta_{\rm A}=0.5$, and 
$\eta_{\rm A}=0.1$. As expected, when the vacuum probability $p_0$ augments, 
the probability $p_{\rm acc}$ decreases and, therefore, the secret key rate decreases as well. 
Similarly,
if the multiphoton probability $p_2$ increases, the intrinsic noise of Alice's signals also 
augments, and both the secret key rate and its cutoff point decreases. 
For comparison purposes, we consider as well the situation where Alice uses an active source 
(thick solid line)
in combination with a BS whose transmittance is optimized with the 
distance. In this last case, the slope of the lower bound on R 
(when Alice employs an active setup)
increases slightly 
when the transmittance of this additional BS starts to be less than one. This occurs, respectively, at
$l\approx{}71$ km (see Fig.~\ref{figure5}), and 
$l\approx{}22$ km (see Fig.~\ref{figure_6b}). From these 
results, we see that 
also in these two examples
the performance of the passive scheme is comparable to the 
one of an active setup when $\eta_{\rm A}$ is sufficiently high. 
Furthermore, it is interesting 
to note that  
the distance covered by a passive transmitter might be even longer than the one of 
an active configuration for some values of the photon number distribution $p_n$ and of the 
parameter $\eta_{\rm A}$ (see, for instance, the dashed line in Fig.~\ref{figure5}). 
Let us emphasize, moreover, that, as discussed above, in our security analysis 
we assume that multiphoton signals are always insecure 
(which is actually true only in the case of an active device), and, therefore, there 
might be still further room for improvement in the secret key rate which can be 
achieved with a passive 
device. 

The passive source shown in Fig.~\ref{figure4} might be 
an interesting alternative to implement the BB84 QKD protocol with 
practical SPS. However, let us mention that to date
there are still two main experimental challenges 
that need to be overcomed in order to obtain relevant secret key rates with 
such a device. 
On the one hand, we need to develop single photon detectors with 
high quantum efficiency that can operate at high clock rate. 
This represents an interesting technological challenge, 
especially at telecom wavelengths where the 
efficiency of present-day single photon detectors lie typically at roughly 10-15$\%$. In recent years, 
however, this field has advanced substantially and there are reasons
to be optimistic; for instance, 
it has been shown lately that  
superconducting transition-edge sensor detectors 
can provide photon number 
resolving capabilities
at telecom wavelengths with 95$\%$ efficiency and negligible noise \cite{super}.
At visible wavelengths, silicon avalanche photodiodes (APD) are already commercially 
available to offer high quantum efficiencies (up to about 70-80$\%$) and low 
dark count rates. This last scenario, for example, is particularly 
relevant for free-space QKD. On the other hand, we need to design
high-quality on-demand SPS. Note that the photon number statistics  
used for simulations purposes 
in this 
section 
assume on-demand SPS that
are still beyond our present 
experimental capability \cite{single_photon}. For instance, the 
normalized second-order correlation function $g^{(2)}$
and 
average photon number per pulse ${\bar n}$ of the 
sources used in Figs.~\ref{figure5} and \ref{figure_6b}
are, respectively, 
$g^{(2)}=0.0039$ and ${\bar n}=0.992$, and $g^{(2)}=0.0452$ and 
${\bar n}=0.815$. 
Despite the remarkable progress that has been made in recent years 
to generate indistinguishable on demand single photons, nowadays 
sub-Poissonian sources have   
a normalized second-order correlation function that lies 
typically between 0.05 and 0.2 \cite{single_photon}. 
However, the biggest 
experimental challenge here is to increase the achievable collection efficiency, which is 
usually below 10$\%$. This means that the probability to obtain
an empty pulse is still rather high, and therefore ${\bar n}$ is 
relatively small \cite{single_photon}. 
As a result, the acceptance probability $p_{\rm acc}$ of the passive BB84 source 
shown in Fig.~\ref{figure4}
would be also quite small.
Alternatively, one may use heralded SPS based on a parametric down 
conversion process, where the emission of an individual photon is 
heralded by the detection of the twin-photon \cite{heralded}.
For instance, the photon number 
distributions of the sources used above can be generated
with this type of sources. In this last case, however, 
the secret 
key rate formula should include 
an additional factor accounting for this detection probability. As a consequence, 
the final key rate can be substantially reduced.  

\begin{figure}
\begin{center}
\includegraphics[angle=0,scale=0.66]{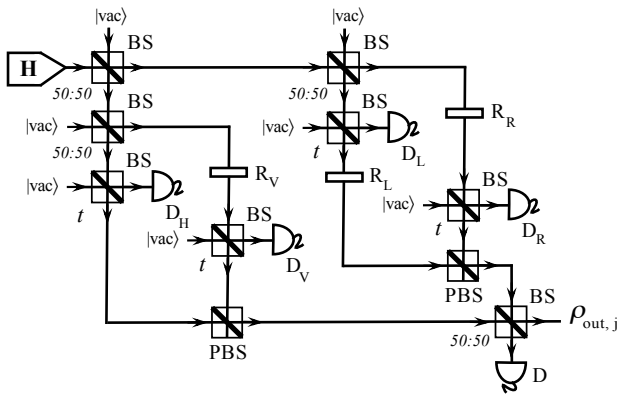}
\end{center}
\caption{Alternative implementation scheme with only one photon source. 
The polarization rotators $R_{\rm V}$, $R_{\rm L}$, and $R_{\rm R}$, 
change the horizontal polarization of the incoming pulses to vertical 
polarization, left circular polarization, and right circular polarization, respectively. 
\label{figure4_b}}
\end{figure}

To conclude this section, let as mention that (as in Sec.~\ref{alter}) 
instead of using the passive source illustrated in Fig.~\ref{figure4}, 
Alice 
could as well employ, for instance, the alternative scheme illustrated in 
Fig.~\ref{figure4_b}.
This setup is similar to the one shown in Fig.~\ref{figure4}, but 
has only one photon source, which might 
make it more robust against side-channel attacks hidden in the imperfections of the 
light sources. 
The main idea is to replace four single-photon pulses (with different polarizations) emitted by four SPS by one 
four-photon pulse (with all their photons prepared in the same polarization state) together 
with polarization rotators. 
The argumentation here goes similar to the passive 
device presented in Fig.~\ref{figure4}, and 
we omit it for simplicity. The resulting secret key rate 
in this scenario, however, might be lower than that in the passive setup analyzed in 
Sec.~\ref{sec_two}, since now the probability $p_{\rm acc}$ to consider an output 
pulse as valid is also lower. Note that the passive scheme shown in Fig.~\ref{figure4_b} has the 
additional requirement that each photon of a four-photon pulse needs to follow a different optical path, which is 
selected by means of $50:50$ BS's. The photon number statistics of the output signals 
generated by both passive schemes are also different.

\section{Conclusion}\label{conc}

Typical experimental realizations of quantum key distribution (QKD) protocols
prepare the signal states by means of an active source. 
In this article, we have investigated two different methods 
to passively generate the signal states of the Bennett-Brassard 1984 (BB84) QKD protocol. Our 
methods need only linear 
optical components and photodetectors, and represent an alternative to those active 
sources that use external-driven elements. 

In particular, 
we have showed that both coherent light and practical single photon sources are 
suitable for passive generation of BB84 signals.
In the asymptotic limit of an infinite long experiment, we have proved that the secret key rate delivered by a
passive source with coherent light is similar to the one provided by an active source, 
thus showing the practical 
interest of the passive scheme.  
When Alice uses 
practical single photon sources, we have showed that 
the distance covered by a passive transmitter might be 
longer 
than the one of an active configuration.
This result is caused by the capacity of the passive scheme to 
reduce the number of multiphoton emissions of the source 
via a post-selection mechanism. 

The main focus of this paper
has been polarization-based realizations of the BB84 protocol, which are particularly relevant for 
free-space QKD. However, we have also showed  
that similar ideas can as well be applied 
to other practical scenarios with
different signal encodings, like, for instance, those QKD experiments based on 
phase encoding, which are 
more suitable to use in combination with optical fibers. 

\section{Acknowledgements}

The authors thank R. Kaltenbaek and B. Qi for very useful discussions. 
M. C. especially thanks the Institute for Quantum Computing 
(University of Waterloo) and the University of Toronto 
for hospitality and support during his stay in 
both institutions. This work was supported by 
CFI, CIPI, the CRC program, CIFAR, MITACS, NSERC, OIT, QuantumWorks, 
and by 
Xunta de Galicia (Spain, grant 
INCITE08PXIB322257PR).

\appendix

\section{Bob's detection setup and channel model}\label{channel}

In this Appendix we include a simple model to characterize Bob's detection device
and the behavior of the quantum channel. This model is used in Sec.~\ref{sec_one} and 
Sec.~\ref{sec_two}
to evaluate the performance of 
the two 
passive BB84 state preparation schemes that we present there. 

For simplicity, in our calculations we shall consider that Bob employs an active BB84 
detection setup. Note, however,  
that the results contained in Sec.~\ref{sec_one} and Sec.~\ref{sec_two} can be straightforwardly 
adapted to cover as well the case where Bob uses a detection apparatus with passive basis 
choice \cite{rarity}. 

Specifically, 
we assume that Bob's detection scheme consists of a polarization analyzer and a 
polarization shifter which effectively changes the polarization basis of the 
subsequent measurement. The polarization analyzer has two detectors, each of them
monitoring the output of a PBS. These detectors are characterized by 
their detection efficiency $\eta_{\rm det}$ and their dark count rate $\epsilon_{\rm B}$ \cite{note2}. 
Furthermore, we suppose that both detectors are equal, and cannot distinguish the 
number of photons of arrival signals. They provide only 
two possible outcomes: ``click" (at least one photon is detected), and ``no click" 
(no photon is detected in the pulse).

The action of such detection device can be described
by two positive operator value measures (POVM), one for 
each of the two BB84 polarization bases $\beta\in\{l,c\}$, 
where $l$ denotes a linear polarization basis and $c$ is a circular 
polarization basis \cite{note1}. 
Each POVM contains four elements: $G_{\rm vac}^\beta$, $G_{\rm 0}^\beta$,
$G_{\rm 1}^\beta$, and $G_{\rm dc}^\beta$. The outcome of the first operator, 
$G_{\rm vac}^\beta$, 
corresponds to no click in the detectors, the following two 
POVM operators, $G_{\rm 0}^\beta$ and
$G_{\rm 1}^\beta$, give precisely one detection 
click (these are the desired measurements), and the last one, 
$G_{\rm dc}^\beta$,
gives rise to both detectors being triggered.
These operators can be written as
\begin{eqnarray}\label{gees}
G_{\rm vac}^\beta&=&[1-\epsilon_{\rm B}(2-\epsilon_{\rm B})]F_{\rm vac}^\beta, \nonumber \\
G_{\rm 0}^\beta&=&(1-\epsilon_{\rm B})\epsilon_{\rm B}F_{\rm vac}^\beta+(1-\epsilon_{\rm B})F_{\rm 0}^\beta, \nonumber \\
G_{\rm 1}^\beta&=&(1-\epsilon_{\rm B})\epsilon_{\rm B}F_{\rm vac}^\beta+(1-\epsilon_{\rm B})F_{\rm 1}^\beta, \nonumber \\
G_{\rm dc}^\beta&=&\openone-G_{\rm vac}^\beta-G_{\rm 0}^\beta-G_{\rm 1}^\beta, 
\end{eqnarray}
where the operators $F_{\rm vac}^\beta$, $F_{\rm 0}^\beta$,
$F_{\rm 1}^\beta$, and $F_{\rm dc}^\beta$ are defined below. Eq.~(\ref{gees}) assumes that 
the background rate, is, to a good approximation, independent of 
the signal detection. Moreover, for easiness of notation, we only consider a background contribution coming from dark counts of Bob's
detectors and we neglect other background 
contributions like, for instance, stray light arising from timing pulses which are not 
completely filtered out in reception. 

The definition 
of the operators  $F_{\rm vac}^\beta$, $F_{\rm 0}^\beta$,
$F_{\rm 1}^\beta$, and $F_{\rm dc}^\beta$
includes as well the effect of the quantum channel. These operators characterize Bob's 
detection device, together with the action of the quantum channel, in the case of noiseless detectors. 
We consider a simple model of a quantum channel
in the absence of eavesdropping; it just consists of a BS of transmittance $\eta_{\rm channel}$. 
For simplicity, here we neglect any misalignment effect in the channel. In this scenario, 
the operators $F_{\rm vac}^\beta$, $F_{\rm 0}^\beta$,
$F_{\rm 1}^\beta$, and $F_{\rm dc}^\beta$
have the form \cite{squash1,mcnl}
\begin{eqnarray}\label{efes}
F_{\rm vac}^\beta&=&\sum_{n,m=0}^\infty (1-\eta_{\rm sys})^{n+m}\ket{n,m}_\beta\bra{n,m}, \nonumber \\
F_{\rm 0}^\beta&=&\sum_{n,m=0}^\infty [1-(1-\eta_{\rm sys})^{n}](1-\eta_{\rm sys})^m\ket{n,m}_\beta\bra{n,m}, \nonumber \\
F_{\rm 1}^\beta&=&\sum_{n,m=0}^\infty [1-(1-\eta_{\rm sys})^{m}](1-\eta_{\rm sys})^n\ket{n,m}_\beta\bra{n,m}, \nonumber \\
F_{\rm dc}^\beta&=&\sum_{n,m=0}^\infty [1-(1-\eta_{\rm sys})^{n}][1-(1-\eta_{\rm sys})^{m}] \nonumber \\
&\times&\ket{n,m}_\beta\bra{n,m}, 
\end{eqnarray}
with $\beta\in\{l,c\}$. The signals 
$\ket{n,m}_l$ ($\ket{n,m}_c$) represent the state which has $n$ photons in the horizontal (circular left) 
polarization mode and $m$ 
photons in the vertical (circular right) polarization mode. The parameter $\eta_{\rm sys}$ denotes the overall 
transmittance of the system. 
This quantity can be written as
\begin{equation}
\eta_{\rm sys}=\eta_{\rm channel}\eta_{\rm B},
\end{equation}
where 
$\eta_{\rm B}$ denotes the overall transmittance of Bob's detection apparatus, {\it i.e.}, $\eta_{\rm B}$ includes 
the transmittance of any optical component within Bob's measurement device together with the 
efficiency $\eta_{\rm det}$ of his detectors. 

The parameter $\eta_{\rm channel}$
can be related with a transmission distance $d$ measured in km for 
the given QKD scheme as 
\begin{equation}
\eta_{\rm channel}=10^{-\frac{\alpha{}d}{10}},
\end{equation}
where $\alpha$ represents 
the loss coefficient of the channel ({\it e.g.}, free-space, or an optical fiber) measured in dB/km. 

\section{Gain and QBER of a passive BB84 QKD setup with coherent light}\label{QE_coherent}

In this Appendix, we calculate the observed gain $Q$ and error rate $E$ for the passive 
QKD device introduced in Sec.~\ref{sec_one}. For that, we use the channel model and 
detection apparatus described in Appendix~\ref{channel}.
In the 
scenario considered, it turns out that the gain is independent of the actual polarization 
of the signals $\rho_{{\rm out},\theta}$ given by Eq.~(\ref{basicstate}) and the basis $\beta$ used to 
measure them. This parameter has the form
\begin{equation}
Q=1-{\rm Tr}(G_{\rm vac}^\beta\rho_{{\rm out},\theta})=1-(1-\epsilon_{\rm B})^2e^{-\mu\eta_{\rm sys}}, 
\end{equation}
for all $\theta$ and $\beta$. 

The calculation of $E$ is slightly more involved, since the error rate
varies depending on the value of the angle $\theta$. By symmetry, however, we can restrict ourselves to investigate 
the QBER in only one of the
valid regions illustrated in Fig.~\ref{figure1_bloch}; note that the error rate is the same in all of them.
For instance, let us consider the case where $\theta\in[7\pi/4+\Omega,\pi/4-\Omega]$ 
(which corresponds to the horizontal polarization interval), and 
let $E_\theta$ denote the error rate of a signal state 
$\rho_{{\rm out}, \theta}$ in that region. This quantity 
can be written as
\begin{equation}\label{errorn}
E_\theta={\rm Tr}(G_{\rm 1}^l\rho_{{\rm out},\theta})+\frac{1}{2}{\rm Tr}(G_{\rm dc}^l\rho_{{\rm out},\theta}).
\end{equation}
The first term in the summation represents the probability that the signal hits the wrong detector on Bob's side, 
{\it i.e.}, he observes a click in the detector associated with vertical polarization. 
The second term in Eq.~(\ref{errorn}) is the probability to have a double click on 
his detection apparatus and assign to it 
a single click in the wrong detector. 
Using Eqs.~(\ref{gees})-(\ref{efes}), we have that 
$E_\theta$ can be further simplified as
\begin{eqnarray}
E_\theta&=&\frac{1}{2Q}\Big\{\epsilon_{\rm B}(\epsilon_{\rm B}-1)f_{{\rm 0},\theta}+\big[2+\epsilon_{\rm B}(\epsilon_{\rm B}-3)\big]f_{{\rm 1},\theta}\nonumber \\
&+&\big[1+\epsilon_{\rm B}(\epsilon_{\rm B}-2)\big]f_{{\rm dc},\theta}+\epsilon_{\rm B}(2-\epsilon_{\rm B})\Big\},
\end{eqnarray}
where 
\begin{equation}
f_{i,\theta}={\rm Tr}(F_i^l\rho_{{\rm out},\theta}),
\end{equation}
for all $i\in\{{\rm 0,1,dc}\}$, and $\theta\in[7\pi/4+\Omega,\pi/4-\Omega]$. 
In order to calculate the probabilities $f_{i,\theta}$ we use Eqs.~(\ref{fock1})-(\ref{fock2})
to first rewrite the state $\rho_{{\rm out},\theta}$ given by Eq.~(\ref{basicstate}) as
\begin{eqnarray}
\rho_{{\rm out}, \theta}&=&e^{-\mu}\sum_{n=0}^\infty \frac{\mu^n}{n!} \frac{1}{4^n}\sum_{k,k'=0}^n \sqrt{\binom{n}{k}\binom{n}{k'}} \nonumber \\
&\times&\big(1+e^{i\theta}\big)^k\big(1+e^{-i\theta}\big)^{k'}\big(1-e^{i\theta}\big)^{n-k} \nonumber \\
&\times&\big(1-e^{-i\theta}\big)^{n-k'}\ket{k,n-k}_{l}\bra{k',n-k'},
\end{eqnarray} 
where $\ket{n,m}_l$ represents again the state which has $n$ photons in the horizontal polarization mode
and $m$ in the vertical polarization mode. We obtain
\begin{eqnarray}
f_{{\rm 0},\theta}&=&e^{-\eta_{\rm sys}\mu}\Big[-1+e^{\frac{1}{2}\eta_{\rm sys}\mu(1+\cos{\theta})}\Big], \nonumber \\
f_{{\rm 1},\theta}&=&e^{-\eta_{\rm sys}\mu}\Big[-1+e^{\frac{1}{2}\eta_{\rm sys}\mu(1-\cos{\theta})}\Big],
\end{eqnarray}
and
\begin{eqnarray}
f_{{\rm dc},\theta}&=&1+e^{-\eta_{\rm sys}\mu}-e^{-\frac{1}{2}\eta_{\rm sys}\mu(1+\cos{\theta})} \nonumber \\
&&-e^{-\eta_{\rm sys}\mu\sin^2{(\frac{\theta}{2})}}.
\end{eqnarray}
The quantum bit error rate $E$ is then given by
\begin{equation}
E=\frac{2}{\pi-4\Omega}\int_{\frac{7\pi}{4}+\Omega}^{\frac{\pi}{4}-\Omega} E_\theta\ \ud \theta,
\end{equation}
and we solve this equation numerically. 

\section{Conditional output state}\label{cond_state}

In this Appendix we provide a mathematical expression for the conditional output state 
$\rho_{{\rm out},j}$ introduced in Sec.~\ref{sec_two} together with its associated probability. 
This signal 
can be written as
\begin{equation}\label{basicstate2}
\rho_{{\rm out},j}=\frac{1}{N}\sum_{n,m,k,l=0}^\infty q_{(n,m,k,l)} \sigma_j^{n,m,k,l},
\end{equation}
where the parameters $q_{(n,m,k,l)}$ are given by 
$q_{(n,m,k,l)}=u_nq_mq_kq_l$,
with
\begin{eqnarray}\label{un}
u_{n}&=&(1-\epsilon_{\rm A})t^n\sum_{k=n}^\infty p_k \binom{k}{k-n}\Big[(1-\eta_{\rm det}) \nonumber \\
&\times&(1-t)\Big]^{k-n}, 
\end{eqnarray}
and
\begin{eqnarray}\label{qn}
q_{n}&=&t^n\sum_{k=n}^\infty p_k \binom{k}{k-n}(1-t)^{k-n}\Big[1-(1-\epsilon_{\rm A}) \nonumber \\
&\times&(1-\eta_{\rm det})^{k-n}\Big].
\end{eqnarray}
The states $\sigma_j^{n,m,k,l}$ have the form
\begin{widetext}
\begin{eqnarray}\label{state_long}
\sigma_j^{n,m,k,l}&=&(1-\epsilon_{\rm A})\frac{2^{-(n+m)}4^{-(k+l)}}{n!m!k!l!}\sum_{w,w'=0}^{k+l}\sum_{r=0}^{{\rm min}\{w+n,w'+n\}}\sum_{s=0}^{{\rm min}\{k+l+m-w,k+l+m-w'\}}
g_{(n,m,k,l,w,w',r,s)}(1-\eta_{\rm det})^{r+s} \nonumber \\
&\times&h_{(n,m,k,l,w,w',r,s)}\ket{w+n-r,k+l+m-w-s}_j\bra{w'+n-r,k+l+m-w'-s},
\end{eqnarray}
\end{widetext}
where $\ket{n,m}_j$ represents the state which has $n$ photons in the polarization mode 
associated with the signal detector $D_j$ which 
did not click, and $m$ 
photons in its orthogonal polarization mode.
The functions $g_{(n,m,k,l,w,w',r,s)}$ and $h_{(n,m,k,l,w,w',r,s)}$ 
which appear in Eq.~(\ref{state_long})
are defined, 
respectively, as
\begin{eqnarray}
g_{(n,m,k,l,w,w',r,s)}&=&f_{(w,k,l)}f_{(w',k,l)}f_{(r,w,n)}f_{(r,w',n)}\nonumber \\
&\times&f_{(s,k+l-w,m)}f_{(s,k+l-w',m)}, 
\end{eqnarray}
with $f_{(x,y,z)}$ given by
\begin{equation}
f_{(x,y,z)}=\sum_{s={\rm max}\{0,x-z\}}^{{\rm min}\{x,y\}} \binom{y}{s}\binom{z}{x-s} (-1)^{z-x+s},
\end{equation}
and
\begin{eqnarray}
&&h_{(n,m,k,l,w,w',r,s)}=r!s!\sqrt{(w+n-r)!}\sqrt{(w'+n-r)!}\nonumber \\
&&\times\sqrt{(k+l+m-w-s)!}\sqrt{(k+l+m-w'-s)!}.
\end{eqnarray}

The normalization factor $N$ of the output states $\rho_{{\rm out},j}$ does not 
depend on the parameter $j$. It can be calculated as 
\begin{equation}\label{N}
N=\sum_{n,m,k,l=0}^\infty q_{(n,m,k,l)} {\rm Tr}(\sigma_j^{n,m,k,l}),
\end{equation}
where the quantities ${\rm Tr}(\sigma_j^{n,m,k,l})$ are given by
\begin{eqnarray}
{\rm Tr}(\sigma_j^{n,m,k,l})&=&(1-\epsilon_{\rm A})\frac{2^{-(n+m)}4^{-(k+l)}}{n!m!k!l!}\sum_{w=0}^{k+l}\sum_{r=0}^{w+n}\nonumber \\
&&\sum_{s=0}^{k+l+m-w}(1-\eta_{\rm det})^{r+s}g_{(n,m,k,l,w,w,r,s)} \nonumber \\
&\times&h_{(n,m,k,l,w,w,r,s)}.
\end{eqnarray}
for all $j$.

Finally, we have that the probability that Alice produces a valid output state, 
{\it i.e.}, only three of her signal detectors $D_i$ click and
$D$ does not click, that we shall denote as $p_{\rm acc}$, is given by
\begin{equation}
p_{\rm acc}=4N.
\end{equation}

\section{Gain, QBER, and multiphoton probability of a passive BB84 QKD setup with practical SPS}\label{sps_par}

It turns out that the gain 
of the protocol is independent of the polarization $j$
of the signals $\rho_{{\rm out},j}$ and the basis $\beta$ used to 
measure them. This parameter can be expressed as
\begin{eqnarray}
Q&=&1-{\rm Tr}(G_{\rm vac}^\beta\rho_{{\rm out},j})=1-\frac{[1-\epsilon_{\rm B}(2-\epsilon_{\rm B})](1-\epsilon_{\rm A})}{N} \nonumber \\
&\times&\sum_{n,m,k,l=0}^\infty q_{(n,m,k,l)}\frac{2^{-(n+m)}4^{-(k+l)}}{n!m!k!l!}\sum_{w=0}^{k+l}\sum_{r=0}^{w+n} \nonumber \\
&\times&\sum_{s=0}^{k+l+m-w}(1-\eta_{\rm det})^{r+s}(1-\eta_{\rm sys})^{n+m+k+l-r-s} \nonumber \\
&\times&g_{(n,m,k,l,w,w,r,s)}h_{(n,m,k,l,w,w,r,s)},  
\end{eqnarray}
for all $j$ and $\beta$.

The QBER has also the same value for all possible polarization states $\rho_{{\rm out},j}$ emitted by Alice. 
Therefore, like in Appendix~\ref{QE_coherent}, in order 
to calculate it we can restrict ourselves to only one given polarization $j$. 
For instance, let us assume that the state produced by the source is $\rho_{{\rm out},{\rm H}}$,
that is, $j={\rm H}$. In 
this case, the error rate can be written as
\begin{equation}
E={\rm Tr}(G_{\rm 1}^l\rho_{{\rm out},{\rm H}})+\frac{1}{2}{\rm Tr}(G_{\rm dc}^l\rho_{{\rm out},{\rm H}}),
\end{equation}
where the first term is the probability that the signal hits the wrong detector on Bob's side, and the 
second term denotes the probability to have a double click and assign to it 
a single click in the wrong detector. Like in Appendix~\ref{QE_coherent}, this quantity can be further simplified 
as
\begin{eqnarray}\label{appp1}
E&=&\frac{1}{2Q}\Big\{\epsilon_{\rm B}(\epsilon_{\rm B}-1)f_{{\rm 0},{\rm H}}+\big[2+\epsilon_{\rm B}(\epsilon_{\rm B}-3)\big]f_{{\rm 1},{\rm H}}\nonumber \\
&+&\big[1+\epsilon_{\rm B}(\epsilon_{\rm B}-2)\big]f_{{\rm dc},{\rm H}}+\epsilon_{\rm B}(2-\epsilon_{\rm B})\Big\},
\end{eqnarray}
where $f_{i,{\rm H}}={\rm Tr}(F_i^l\rho_{{\rm out},{\rm H}})$ for all $i\in\{{\rm 0,1,dc}\}$. 
The probabilities $f_{i,{\rm H}}$ can be obtained directly using Eqs.~(\ref{efes})-(\ref{basicstate2}).
In particular, 
we find that
\begin{eqnarray}
f_{{\rm 0},{\rm H}}&=&\frac{(1-\epsilon_{\rm A})}{N} \sum_{n,m,k,l=0}^\infty q_{(n,m,k,l)}\frac{2^{-(n+m)}4^{-(k+l)}}{n!m!k!l!}\nonumber \\
&\times&\sum_{w=0}^{k+l}\sum_{r=0}^{w+n}\sum_{s=0}^{k+l+m-w}(1-\eta_{\rm det})^{r+s}\nonumber \\
&\times&[1-(1-\eta_{\rm sys})^{w+n-r}](1-\eta_{\rm sys})^{k+l+m-w-s}\nonumber \\
&\times&g_{(n,m,k,l,w,w,r,s)}h_{(n,m,k,l,w,w,r,s)},\nonumber \\
f_{{\rm 1},{\rm H}}&=&\frac{(1-\epsilon_{\rm A})}{N} \sum_{n,m,k,l=0}^\infty q_{(n,m,k,l)}\frac{2^{-(n+m)}4^{-(k+l)}}{n!m!k!l!}\nonumber \\
&\times&\sum_{w=0}^{k+l}\sum_{r=0}^{w+n}\sum_{s=0}^{k+l+m-w}(1-\eta_{\rm det})^{r+s}(1-\eta_{\rm sys})^{w+n-r} \nonumber \\
&\times&[1-(1-\eta_{\rm sys})^{k+l+m-w-s}]g_{(n,m,k,l,w,w,r,s)} \nonumber \\
&\times&h_{(n,m,k,l,w,w,r,s)}, 
\end{eqnarray}
and
\begin{eqnarray}
f_{{\rm dc},{\rm H}}&=&\frac{(1-\epsilon_{\rm A})}{N} \sum_{n,m,k,l=0}^\infty q_{(n,m,k,l)}\frac{2^{-(n+m)}4^{-(k+l)}}{n!m!k!l!}\nonumber \\
&\times&\sum_{w=0}^{k+l}\sum_{r=0}^{w+n}\sum_{s=0}^{k+l+m-w}(1-\eta_{\rm det})^{r+s}\nonumber \\
&\times&[1-(1-\eta_{\rm sys})^{w+n-r}][1-(1-\eta_{\rm sys})^{k+l+m-w-s}]\nonumber \\
&\times&g_{(n,m,k,l,w,w,r,s)}h_{(n,m,k,l,w,w,r,s)}.
\end{eqnarray}

Finally, the multiphoton probability of the source, $p_{\rm multi}$, is as well independent 
of the value of the polarization $j$. It can be obtained as
\begin{eqnarray}\label{appp2}
p_{\rm multi}&=&1-\sum_{n=0}^{1}\sum_{m=0}^{1-n} {\rm Tr}\Big(\ket{n,m}_j\bra{n,m}\rho_{{\rm out},j}\Big) \nonumber \\
&=&1-p_{0,0,j}-p_{0,1,j}-p_{1,0,j}, 
\end{eqnarray}
where the probabilities $p_{n,m,j}$ are defined as
\begin{equation}
p_{n,m,j}\equiv\bra{n,m}\rho_{{\rm out},j}\ket{n,m}_j.
\end{equation} 
After a short calculation, we find that
\begin{eqnarray}
p_{0,0,j}&=&\frac{1}{N} \sum_{n,m,k,l=0}^\infty q_{(n,m,k,l)}\frac{2^{-(n+m)}4^{-(k+l)}}{n!m!k!l!}\nonumber \\
&\times&(1-\epsilon_{\rm A})\sum_{w=0}^{k+l}(1-\eta_{\rm det})^{n+m+k+l}\nonumber \\
&\times&g_{(n,m,k,l,w,w,w+n,k+l+m-w)}\nonumber \\
&\times&h_{(n,m,k,l,w,w,w+n,k+l+m-w)},
\end{eqnarray}
\begin{eqnarray}
p_{0,1,j}&=&\frac{1}{N} \sum_{n,m,k,l=0}^\infty q_{(n,m,k,l)}\frac{2^{-(n+m)}4^{-(k+l)}}{n!m!k!l!}\nonumber \\
&\times&\sum_{w=0}^{{\rm min}\{k+l,k+l+m-1\}}(1-\eta_{\rm det})^{n+m+k+l-1}\nonumber \\
&\times&(1-\epsilon_{\rm A})g_{(n,m,k,l,w,w,w+n,k+l+m-w-1)}\nonumber \\
&\times&h_{(n,m,k,l,w,w,w+n,k+l+m-w-1)}, 
\end{eqnarray}
and
\begin{eqnarray}
p_{1,0,j}&=&\frac{1}{N} \sum_{n,m,k,l=0}^\infty q_{(n,m,k,l)}\frac{2^{-(n+m)}4^{-(k+l)}}{n!m!k!l!}\nonumber \\
&\times&\sum_{w={\rm max}\{0,1-n\}}^{k+l}(1-\eta_{\rm det})^{n+m+k+l-1}\nonumber \\
&\times&(1-\epsilon_{\rm A})g_{(n,m,k,l,w,w,w+n-1,k+l+m-w)}\nonumber \\
&\times&h_{(n,m,k,l,w,w,w+n-1,k+l+m-w)}.
\end{eqnarray}

\bibliographystyle{apsrev}

\begin{thebibliography}{99}

\bibitem{qkd} 
N. Gisin, G. Ribordy, W. Tittel and H. Zbinden, Rev. Mod. Phys. {\bf 74}, 145 (2002); 
M. Du\v{s}ek, N. L\"utkenhaus and M. Hendrych, Progress in Optics {\bf 49}, Edt. E.
Wolf (Elsevier), 381 (2006); 
V. Scarani, H. Bechmann-Pasquinucci, N. J. 
Cerf, M. Du\v{s}ek, N. L\"utkenhaus and M. Peev, 
Rev. Mod. Phys. {\bf 81}, 1301 (2009).
\bibitem{Vernam} G. S. Vernam, J. Am. Inst. Electr. Eng. {\bf XLV}, 109 (1926).
\bibitem{bb84} C. H. Bennett and G. Brassard, Proc. IEEE Int.
Conference on Computers, Systems and Signal Processing, Bangalore,
India, IEEE Press, New York, 175 (1984).
\bibitem{4lasers}
R. J. Hughes, J. E. Nordholt, D. Derkacs and C. G. Peterson, New J. Phys. 
{\bf 4}, 43 (2002);
C. Kurtsiefer, P. Zarda, M. Halder, H. Weinfurter, P. M. Gorman, P. R. Tapster and
J. G. Rarity, Nature {\bf 419}, 450 (2002).
\bibitem{modulator}
C. H. Bennett, F. Bessette, G. Brassard, L. Salvail and J. Smolin, J. Cryptology {\bf 5}, 3 (1992);
A. Muller, J. Br\'eguet and N. Gisin, Europhysics Lett. {\bf 23}, 383 (1993);
J. Br\'eguet, A. Muller and N. Gisin, J. Mod. Opt. {\bf 41}, 2405 (1994);
A. Muller, H. Zbinden and N. Gisin, Nature {\bf 378}, 449 (1995); 
A. Muller, H. Zbinden and N. Gisin, Europhysics Lett. {\bf 33}, 335 (1996);
P. Townsend, IEEE Photonics Tech. Lett. {\bf 10}, 1048 (1998); 
G. B. Xavier, N. Walenta, G. Vilela de Faria, G. P. Tempor\~ao, N. Gisin, H. Zbinden and J. P. 
von der Weid, New J. Phys. {\bf 11}, 045015 (2009).
\bibitem{post} N. L\"utkenhaus, Appl. Phys. B: Lasers Opt. {\bf 69}, 395 (1999);
D. Gottesman and H.-K. Lo, IEEE Trans. Inf. Theory {\bf 49}, 457 (2003);
X. Ma, C.-H. F. Fung, F. Dupuis, K. Chen, K. Tamaki and H.-K. Lo, Phys. Rev. A {\bf 74}, 032330 (2006);
V. Scarani, A. Ac\'{\i}n, G. Ribordy and N. Gisin, Phys. Rev. Lett. {\bf 92}, 057901 (2004);
B. Kraus, N. Gisin and R. Renner, Phys. Rev. Lett. {\bf 95}, 080501 (2005);
R. Renner, N. Gisin and B. Kraus, Phys. Rev. A {\bf 72}, 012332 (2005);
J. M. Renes and Graeme Smith, Phys. Rev. Lett. {\bf 98}, 020502 (2007).
\bibitem{sec_bb84}
H. Inamori, N. L\"utkenhaus and D. Mayers, Eur. Phys. J. D {\bf 41}, 599 (2007).
\bibitem{sec_bb84b}
D. Gottesman, H.-K. Lo, N. L\"utkenhaus and J. Preskill, Quantum Inf. Comput. {\bf 4}, 
325 (2004). 
\bibitem{decoy} W.-Y. Hwang, Phys. Rev. Lett. {\bf 91}, 057901 (2003); 
H.-K. Lo, X. Ma and K. Chen, Phys. Rev. Lett. {\bf 94}, 230504
(2005); X.-B. Wang, Phys. Rev. Lett. {\bf 94}, 230503 (2005);
X.-B. Wang, Phys. Rev. A {\bf 72}, 012322 (2005); 
X.-B. Wang, Phys. Rev. A {\bf 72}, 049908(E) (2005).
\bibitem{decoy_e} 
Y. Zhao, B. Qi, X. Ma, H.-K. Lo and L. Qian, Phys. Rev. Lett.
{\bf 96}, 070502 (2006); 
Y. Zhao, B. Qi, X. Ma, H.-K. Lo and L.
Qian, Proc. of IEEE International Symposium on Information Theory
(ISIT'06), 2094 (2006); 
C.-Z. Peng, J. Zhang, D. Yang, W.-B. Gao,
H.-X. Ma, H. Yin, H.-P. Zeng, T. Yang, X.-B. Wang and J.-W. Pan,
Phys. Rev. Lett. {\bf 98}, 010505 (2007); 
D. Rosenberg, J. W.
Harrington, P. R. Rice, P. A. Hiskett, C. G. Peterson, R. J.
Hughes, A. E. Lita, S. W. Nam and J. E. Nordholt, Phys. Rev.
Lett. {\bf 98}, 010503 (2007); T. Schmitt-Manderbach, H. Weier,
M. F\"urst, R. Ursin, F. Tiefenbacher, T. Scheidl, J. Perdigues,
Z. Sodnik, C. Kurtsiefer, J. G. Rarity, A. Zeilinger and H.
Weinfurter, Phys. Rev. Lett. {\bf 98}, 010504 (2007); 
Z. L. Yuan,
A. W. Sharpe and A. J. Shields, Appl. Phys. Lett. {\bf 90}, 011118
(2007); 
Z.-Q. Yin, Z.-F. Han, W. Chen, F.-X. Xu, Q.-L. Wu and
G.-C. Guo, Chin. Phys. Lett {\bf 25}, 3547 (2008); J. Hasegawa, M. Hayashi, T.
Hiroshima, A. Tanaka and A. Tomita, Preprint quant-ph/0705.3081; J. F.
Dynes, Z. L. Yuan, A. W. Sharpe and A. J. Shields, Optics Express
{\bf 15}, 8465 (2007).
\bibitem{random}
J. F. Dynes, Z. L. Yuan, A. W. Sharpe and A. J. Shields, Appl. Phys. Lett. {\bf 93}, 1 (2008); 
B. Qi, Y.-M. Chi, H.-K. Lo and L. Qian, Opt. Lett. {\bf 35}, 312 (2010);
K. Hirano, T. Yamazaki, S. Morikatsu, H. Okumura, H. Aida, A. Uchida, S. Yoshimori, K. Yoshimura, 
T. Harayama and P. Davis, Opt. Express {\bf 18}, 5512 (2010);
M. F\"urst, H. Weier, S. Nauerth, D. G. Marangon, C. Kurtsiefer and H. Weinfurter,
Opt. Express {\bf 18}, 13029 (2010).
\bibitem{rarity} J. G. Rarity, P. C. M. Owens and P. R. Tapster, J. Mod. Opt. {\bf 41}, 2435 (1994).
\bibitem{passive1}
W. Mauerer and C. Silberhorn, Phys. Rev. A {\bf 75}, 050305(R) 
(2007); Y. Adachi, T. Yamamoto, M. Koashi and N. Imoto, Phys. Rev. 
Lett. {\bf 99}, 180503 (2007); 
X. Ma and H.-K. Lo, New J. Phys. {\bf 10}, 073018 (2008).
\bibitem{passive2}
M. Curty, T. Moroder, X. Ma and N. L\"utkenhaus, Opt. Lett. {\bf 34}, 
3238 (2009);
Y. Adachi, T. Yamamoto, M. Koashi and N. Imoto, New J. Phys. {\bf 11}, 113033 (2009);
M. Curty, X. Ma, B. Qi and T. Moroder, Phys. Rev. A {\bf 81}, 022310 (2010). 
\bibitem{passiveEB} 
G. Ribordy, J. Brendel, J.-D. Gautier, N. Gisin and H. Zbinden, Phys. Rev. A {\bf 63}, 012309 (2000);
W. Tittel, J. Brendel, H. Zbinden and N. Gisin, Phys. Rev. Lett. {\bf 84}, 4737 (2000).
\bibitem{koashi_p} M. Koashi and J. Preskill, Phys. Rev. Lett. {\bf 90}, 057902 (2003).
\bibitem{oneway} T. Moroder, M. Curty and N. L\"utkenhaus, Phys. Rev. A {\bf 74}, 052301 (2006).
\bibitem{squash1} N. L\"utkenhaus, Phys. Rev. A {\bf 59}, 3301 (1999).
\bibitem{squash2}
N. L\"utkenhaus, Phys. Rev. A {\bf 61}, 052304 (2000);
T. Tsurumaru and K. Tamaki, Phys. Rev. A {\bf 78}, 032302 (2008); 
N. J. Beaudry, T. Moroder and N. L\"utkenhaus, Phys. Rev. 
Lett. {\bf 101}, 093601 (2008).
\bibitem{eff_bb84} H.-K. Lo, H. F. C. Chau and M. Ardehali, J. Cryptology {\bf 18}, 133 (2005). 
\bibitem{eff_error} G. Brassard and L. Salvail, in {\it Advances in Cryptology EUROCRYPT'93}, edited 
by T. Helleseth (Springer, Berlin), Lecture Notes in Computer Science Vol. {\bf 765}, 410 (1994).  
\bibitem{pump}
P. Kok and S. L. Braunstein, Phys. Rev. A {\bf 61}, 042304 (2000).
\bibitem{pns} B. Huttner, N. Imoto, N. Gisin
and T. Mor, Phys. Rev. A {\bf 51}, 1863 (1995); 
G. Brassard, N. L\"utkenhaus, T. Mor and 
B. C. Sanders, Phys. Rev. Lett. {\bf 85}, 1330 (2000).
\bibitem{super} A. E. Lita, A. J. Miller and S. W. Nam, Opt. Express {\bf 16}, 3032 (2008).
\bibitem{single_photon} B. Lounis and M. Orrit, Rep. Prog. Phys. {\bf 68}, 1129 (2005);
A. J. Shields, Nature Phot. {\bf 1}, 215 (2007);
D. J. P. Ellis, A. J. Bennett, A. J. Shields, P. Atkinson and D. A. Ritchie, Appl. Phys. Lett. {\bf 90}, 233514 (2007);
M. Hijlkema, B. Weber, H. P. Specht, S. C. Webster, A. Kuhn and G. Rempe, Nature Phys. {\bf 3}, 253 (2007).
J. Bochmann, M. M\"ucke, G. Langfahl-Klabes, C. Erbel, B. Weber, H. P. Specht, D. L. Moehring and G. Rempe, 
Phys. Rev. Lett. {\bf 101}, 223601 (2008);
D. A. Simpson, E. Ampem-Lassen, B. C. Gibson, S. Trpkovski, F. M. Hossain, S. T. Huntington, A. D. Greentree, L. C. L. 
Hollenberg and S. Prawer, Appl. Phys. Lett. {\bf 94}, 203107 (2009);
E. B. Flagg, A. Muller, J. W. Robertson, S. Fouta, D. G. Deppe, M. Xiao, W. Ma, G. J. Salamo and 
C. K. Shih, Nature Phys. {\bf 5}, 203 (2009);
D. Englund, B. Shields, K. Rivoire, F. Hatami, 
J. Vu\v ckovi\'c, H. Park and M. D. Lukin, Preprint arXiv:1005.2204.
\bibitem{heralded} T. B. Pittman, B. C. Jacobs and J. D. Franson, Opt. Comm. {\bf 246}, 545 (2004).
\bibitem{note2}
Here we consider that dark counts in the detectors 
are, to a good approximation, independent of the incoming signals. 
\bibitem{note1}  Alternatively, one can also characterize Bob's detection setup 
using only one POVM which already includes the action of the 
polarization shifter.
\bibitem{mcnl} M. Curty and N. L\"utkenhaus, Phys. Rev. A {\bf 69}, 042321 (2004).

\end{thebibliography}

\end{document}